\RequirePackage{fix-cm}
\documentclass[twocolumn,epjc3]{svjour3}  
\smartqed  
\RequirePackage{graphicx}
\journalname{Eur. Phys. J. C}
\begin{document}

\title{The carbon star mystery: forty years later
}
\subtitle{Theory and observations}


\author{Oscar Straniero \thanksref{e1,addr1}
        \and
        Carlos Abia\thanksref{e2,addr2} 
       \and 
      Inma Dom\'\i nguez\thanksref{e3,addr2}
}

\thankstext{e1}{e-mail:oscar.straniero@inaf.it}
\thankstext{e2}{cabia@ugr.es}
\thankstext{e3}{inma@ugr.es}

\institute{INAF - Osservatorio Astronomico d’Abruzzo, Via Maggini snc, I-64100 Teramo, Italy \label{addr1}
           \and
           Dpto. F\'\i sica Te\'orica y del Cosmos, Universidad de Granada, E-18071 Granada, Spain \label{addr2}
}

\date{Received: date / Accepted: date}

\maketitle

\begin{abstract}
In 1981 Icko Iben Jr published a paper entitled ''The carbon star mystery: why do the low mass ones become such, and where have all the high mass ones gone?", where he discussed the discrepancy between the theoretical expectation and its observational counterpart about the luminosity function of AGB carbon stars. After more than 40 years, our understanding of this longstanding problem is greatly improved, also thanks to more refined stellar models and a growing amount of observational constraints. In this paper we review the state of the art of these studies  and we briefly illustrate the future perspectives.    
\keywords{Stellar evolution\and Nucleosynthesis \and Stellar abundances}
\end{abstract}

\section{Introduction}
\label{intro}
Carbon stars (or C stars) are those showing an abundance ratio by number C/O$>1$. They were firstly identified as a separate spectroscopic class by father Angelo Secchi. In a report to the French Academy of Sciences dated back to 1868, he wrote: {\it"...stars which do not belong to the three established types are very rare... I believe that they will belong to the family of the red stars and of variable stars"}. For a long time, the origin of this peculiar stellar chemistry has been a mystery \cite{ibe81}. A carbon enhancement can be produced either by an internal mixing of freshly synthesised carbon (intrinsic carbon stars), or through the accretion of carbon-rich matter from a companion star in a binary system (extrinsic carbon stars). As a consequence, the carbon enhancement may appear at different stages
in the evolution of the stars, and, hence, a variety of carbon star spectral types are possible, depending on the actual effective temperature and gravity, and on the abundances of the molecules bearing carbon atoms (CN, C$_2$, CH, see \cite{wal98}, \cite{abi03}). In this paper we will focus on the so-named normal (N-type) carbon stars, i.e. the intrinsic carbon stars that form during the asymptotic giant branch (AGB) phase. In general, these C stars are believed to be important contributors to the carbon budget in the Galaxy. However, their net contribution is not yet clear, also in comparisons with other possible C polluters, such as 
massive Wolf-Rayet stars. To settle the question and quantify the relative contribution of AGB stars to the evolution of the galactic carbon abundance, it is necessary to know 
how much carbon is produced and ejected as a function of the initial mass and metallicity. The theory of stellar evolution teaches us that surface carbon enrichment in the AGB phase is a consequence of periodic episodes of convective mixing, named the third dredge-up (TDU), which transport material that has suffered He-burning up to the stellar surface \cite{str06}, \cite{kar14}. In practice, an AGB star undergoes recurrent thin-shell instabilities \cite{SH1967}, called thermal pulses (TP), which induce thermonuclear runaways, or He-shell flashes,  whose power may attain a few $10^8$ L$_\odot$ (in AGB stars with M$\sim 2$ M$_\odot$). A TDU episode may occur after a TP, when the external layers expand and cool down, until the H-burning shell eventually dies out, and the external convection can penetrate the He- and C-rich mantle. However, more than 40 years after the pioneering Iben's paper on {\it The Carbon star mystery} \cite{ibe81}, the efficiency of the TDU and the chemical yields from AGB stars are still burdened by heavy uncertainties and disagreements among different authors, mainly due to the lack of a robust theory of convection and mass loss. An homogeneous and accurate set of spectroscopic and photometric observations could compensate such a theoretical drawback. \\
Carbon stars are also among the main sites where  heavy elements (A$\geq 90$) are produced trough the slow capture of neutrons: the s process. Neutrons for this process are provided by the $^{13}$C$(\alpha, n)^{16}$O reaction, which is active  at relatively low temperature ($T\sim 90$ MK) during the period of time that elapses between two TPs (inter-pulse phase). According to the current paradigm, the partial mixing occurring at the bottom of the convective envelope at the time of the TDU leaves a thin pocket where the H mass fraction is $X_H<0.01$, while the carbon mass fraction is about 0.2. Then, at the H re-ignition, a substantial amount of $^{13}$C is produced  by the $^{12}$C$(p, \gamma)^{13}$N reaction followed by the $^{13}$N decay, and, later on, the s-process can start, due to the activation of the  $^{13}$C neutron source. A second neutron burst, as due to the activation of the $^{22}$Ne$(\alpha, n)^{25}$Mg reaction, may eventually occur at the bottom of the convective shell powered by a TP, but only if the temperature exceeds
$300$ MK (see e.g. \cite{kap11}, and
references therein). As a matter of fact, the s-process enhancement observed in normal C stars is mainly due to the $^{13}$C neutron burst, while the $^{22}$Ne source only provides a marginal contribution. The presence of Tc alive ($^{99}$Tc half-life $2.11\times 10^5$ yr) in the atmosphere of C stars is a probe of their intrinsic nature. \\
Carbon stars are also the parents of main stream SiC grains that may form in their cool and C-rich circumstellar envelopes. Some meteorites that hit the Earth contain these stardust grains, which are isolated and analysed in the laboratory. In this way, SiC grains provide valuable information on the physical conditions occurring in the circumstellar envelopes of C stars as well as on the internal nucleosynthesis processes \cite{dav11}. \\
During the last few years a number
of theoretical and observational studies shed
new light on the scenario described above. Here we discuss some of these advances, as obtained by combining new theoretical models and more accurate observational constraints. In section 2 we illustrate state-of-the-art 
models of C-star progenitors and their nucleosynthesis. In Section 3 we discuss recent observational advances and the new issues that these observations have revealed, and in Section 4, we summarise the current status and future prospects of this subject.

\section{Theory \& models}
As discussed by I. Iben in his a seminal pape \cite{ibe81}, the C-star luminosity function (N-type) of the Milky Way and of the Magellanic Clouds are peaked at M$_{\rm{{bol}}}\sim -5$, and very few  C stars are brighter than M$_{\rm{{bol}}}\sim -6$ (see Sect. 3). This implies that i) the majority of the C stars should have mass between 1.5 and 2.5 M$_\odot$, and ii) very rare C stars are observed whose mass exceeds 3-4  M$_\odot$. Since 1981,  many progresses have been done in modelling AGB stars and an answer to these questions have been partially found. Nevertheless, a general consensus has not yet been reached, because of the many uncertainties still affecting AGB stellar models. First of all, let us discuss the widely accepted scenario for the C-star formation.\\
As it is well known, a TP-AGB stars is made of three zones, namely: a C-O-rich core, sustained by the pressure of degenerate electrons and cooled by the release of plasma neutrinos; an intermediate He-rich region, where  recurrent TPs powered by He burning take place; and a H-rich envelope efficiently mixed by a rather deep convective envelope. For most of the time, the He-burning shell is off, and the luminosity is powered by the CNO bi-cycle active in a thin H-burning shell. The compression and heating of the matter left behind by H burning causes the He ignition and a convective shell, which extends over almost the entire He-rich zone, develops. In this way, the carbon produced by the triple-$\alpha$ reaction is mixed-up to the top of the He-rich region. At the end of the TP, the carbon mass fraction in the most external layers of this intermediate zone is raised up to $\sim 20$~\%. So far, the presence of an active H-burning shell has maintained an entropy barrier that prevents the penetration of the external convection into the underlying H-exhausted region. However, due to the outgoing energy flow generated by He burning, the envelope expands and cools, until H burning dies out. In this condition, the external convection can penetrate the H-He discontinuity and, eventually, can reach the C-enhanced zone. In low-mass AGB stars, this deep mixing episode may occur a few hundred years after the TP quenching, while in a massive AGB star it requires a much shorter time. Anyway, the resulting carbon dredge-up is the process responsible for the formation of an intrinsic C star. Depending on the initial metallicity, several TDU episodes may be required until the C/O$>1$ condition is attained at the stellar surface.  So, the question is: in which stars are the TDUs sufficiently intense to allow them to become C stars? In this context, we have understood that the efficiency of the dredge-up process depends on the concurrent actions of several stellar parameters, such as the core and the envelope masses, as well as the  initial metallicity (see, e.g., \cite{straniero2003}). In the following we try to disentangle the various physical processes that affect the carbon dredge-up in AGB stars.  

\subsection{The shell H-burning}
The TDU is the result of the expansion and cooling of the envelope that follows the violent He-ignition. This occurrence causes the temporary stop of H burning and an increase of the radiative opacity, and both of these phenomena favours the penetration of the external convective zone into the H-exhausted region. It goes without saying that the ultimate engine of the TDU is the He-burning thermonuclear runaway. As a matter of fact, deeper TDUs are found after stronger thermal pulses\footnote{The strength of a thermal pulse may be measured by the maximum luminosity attained by He-burning during a TP.}. \\
On the other hand, the H-burning rate during the inter-pulse period determines the He-ignition conditions and, in turn, the strength of the TP. In particular, the He-ignition density is larger in case of slower H burning and, in turn, a higher peak luminosity is attained during the thermal runaway. For instance,  \cite{straniero2000} (see also \cite{herwig2004}) have shown how a reduction of the $^{14}$N$(p,\gamma)^{15}$O reaction rate leads to stronger TPs and deeper TDU episodes. Indeed, this reaction is the bottleneck of the CNO and its rate controls the rate of the shell H burning. On the other hand, the H-burning efficiency also depends on the mass of the H-exhausted core \cite{iben1983} \cite{straniero2003}. Hence, stronger TPs and, in turn, deeper TDUs are found in low-mass AGB stars, those with a smaller core mass and, in turn, a less efficient shell H burning.  
In principle, also the initial metallicity, more precisely the initial abundances of C, N, and O, affects the strength of the first few TPs: the lower the CNO abundance the stronger the thermal pulses and the deeper the TDU. However, in the late part of the AGB, the CNO abundances in the envelope are modified by the TDUs and the influence of the initial metallicity disappears. 
\subsection{The hot-bottom-burning (massive AGB stars only)}
During the inter-pulse periods of massive AGB stars (M $\geq 4$ M$_\odot$), the bottom of the convective envelope penetrates the zone where H burning is active. This phenomenon, which makes massive AGB stars important sites for the nucleosynthesis of various light and intermediate mass isotopes,  is known as  hot-bottom-burning (HBB). The temperature of the deeper layer of the convective envelope may be $\sim30\times 10^6$ K in a 4 M$_\odot$ (solar composition) and up to $100\times 10^6$ K in a 7 M$_\odot$ star. A lower metallicity favour the HBB, because of the less steep entropy barrier at the H-burning shell, which is, for this reason,  more easily penetrated by convective instability. This phenomenon has two major consequences. Firstly, fresh H is brought into the H-burning layers. As a result, the shell H-burning is more efficient and, in turn, the TPs are weaker and the TDUs are shallower. In addition, carbon, primordial or carried in the envelope by the TDU, is mostly transformed into nitrogen through the CN cycle active at the bottom of the external convective region. So, massive AGB stars are expected to become N-rich instead of C-rich.  This process eventually ceases when the envelope mass, which is eroded by mass loss, reduces down to $\sim2.2$ M$_\odot$. Therefore, approaching the AGB tip, a star with mass larger than 4 M$_\odot$ may still become C-rich, but just for a short time.  
\subsection{The hot third dredge-up (massive AGB stars only)}
As previously said, the TDU starts just after a TP, when H burning dies out. In massive AGB stars, however, when the convective instability penetrates the H-exhausted core, it encounters layers where the temperature is sufficiently high to re-activate proton-capture reactions. Then, the energy released by nuclear reactions contrasts the convective instability that is pushed outward, thus causing  a premature stop of the TDU. This phenomenon, called hot third dredge-up (HTDU) significantly limits the TDU in massive AGB stars \cite{goriely2004}. In passing, let us note that  owing to the HTDU, the s-process yields from the more massive AGB stars are expected to be negligible. 
\subsection{The mass-loss}
 The AGB phase terminates when the mass-loss erodes the envelope until H burning is substantially suppressed. For a 2 M$_\odot$ (solar composition), the star is expected to leave the AGB when the envelope mass is reduced down to $\sim0.1$ M$_\odot$, but the TDUs become progressively shallower when the envelope mass becomes lower than $\sim0.5$ M$_\odot$ (see, e.g., \cite{cristallo2009}. In this context, the AGB mass-loss rate determines the number of TDU episodes and the total amount of carbon that is dredged-up during the AGB phase. In other words, the mass-loss rate determines the possibility for an AGB star of becoming a C star or not. In addition, for stars with mass lower than $\sim 2$ M$_\odot$ also the pre-AGB mass-loss play an important role. When these stars leave the main sequence, they enter the RGB phase, during which they lose up to a few tenths of solar masses. Then, these stars approach the TP-AGB phase with an already eroded envelope.
 \subsection{Boundary mixing and extra-mixing}
\begin{figure}
\begin{center}
\includegraphics[scale=0.45]{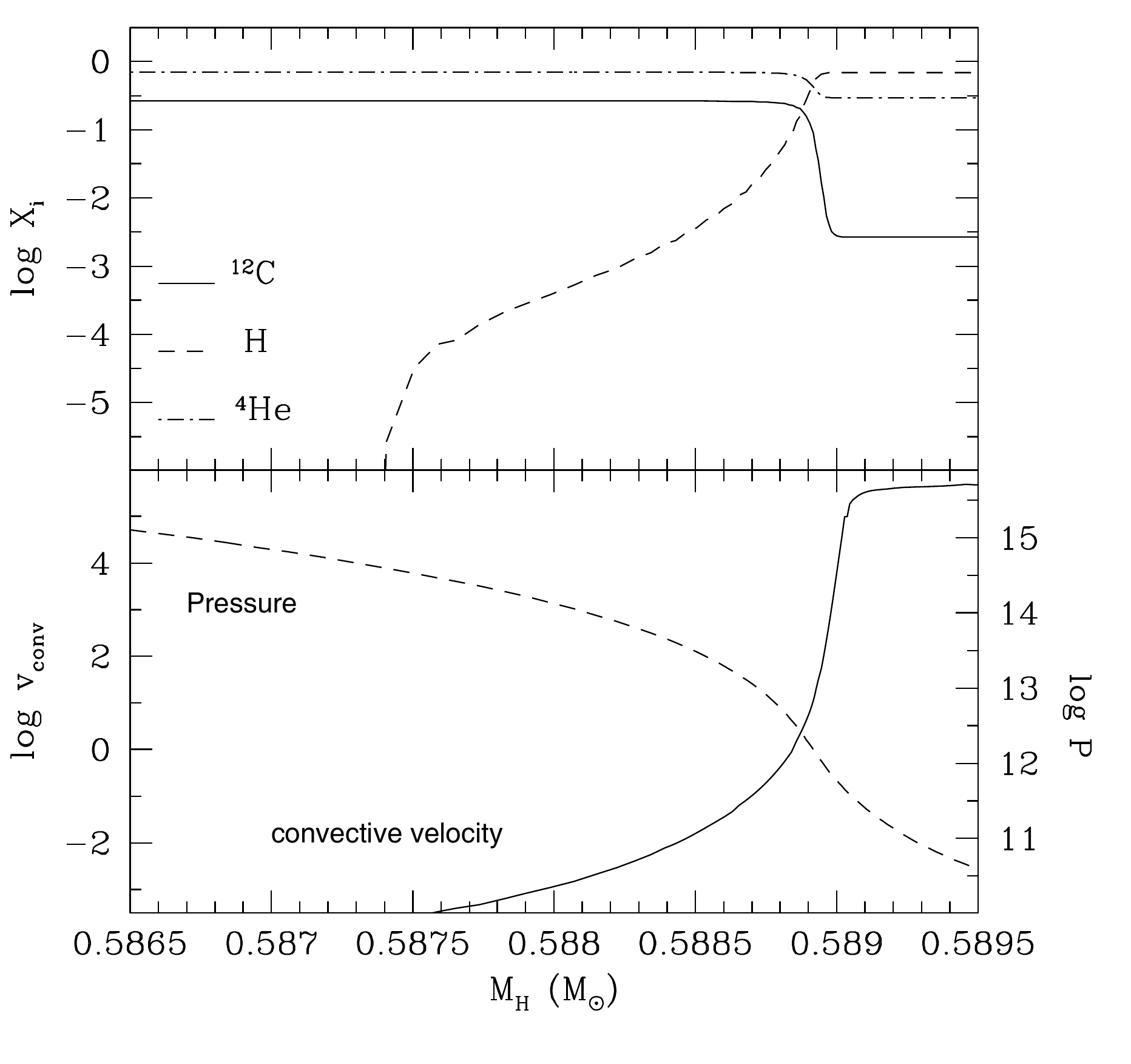}
\caption{The boundary of the convective envelope during the TDU for a 2 M$_\odot$ with solar composition. Upper panel: chemical composition in the transition region between the convective envelope 
and the radiative He-rich zone.
Lower panel: the exponential decline of the convective velocity and the pressure gradient. Adapted from \cite{straniero2006}.
}\label{velco}
\end{center}
\end{figure}

When the convective envelope penetrates the H-exhausted zone, a sharp variation of the composition
takes place at the convective boundary. In less than $10^{-3}$ M$_\odot$, the H mass-fraction drops from about 0.7 to 0. Owing to this composition discontinuity, a sharp variation
of the radiative opacity, associated to an abrupt change of the radiative gradient, develops. In these conditions, the precise location of the convective border (i.e. the limit of the region
fully mixed by convection) becomes highly uncertain. An initially
small perturbation causing a mixing just below the convective boundary  is amplified on a dynamical timescale, so that the radiative gradient in the radiative stable zone rises up and the convective instability moves inward. This condition is commonly
encountered in stellar model computations at the time of the second and the third dredge-up (see e.g., \cite{bec79}, \cite{cas90}, \cite{fro96}, \cite{cas98}, \cite{cristallo2009}). While the effect
of such an instability is marginal in the case of the second dredge-up \cite{cas98}, the deepness of the TDU is significantly extended \cite{cristallo2009}.
In order to correctly treat this phenomenon, a more realistic description of the convective boundary
than that usually adopted in extant stellar evolution codes is required. Instead of a well defined spherical
surface, as obtained when the bare Schwarzschild’s criterion
is used, the transition between the full-radiative core
(i.e. unmixed) and the full-convective (i.e. fully mixed)
envelope likely occurs in an extended zone where
only a partial mixing takes place (semi-convective layer), so that a smooth and stable H-profile may form (see Figure \ref{velco}).  Within this transition zone, the convective velocity smoothly drops from about $10^5$ cm/s, at the convective boundary, to 0. Note that this process may also solve another longstanding issue of AGB stars, that is the formation of the $^{13}$C-pocket needed to activate a substantial s-process nucleosynthesis during the inter-pulse phase (see, e.g., \cite{cristallo2009}). It is indeed in this transition zone left after a TDU episode that a suitable amount of $^{13}$C may form, through the $^{12}$C$(p.\gamma)^{13}$N reaction followed by the $^{13}$N decay.  
In spite of the many efforts done to incorporate this phenomenon in one-dimension hydrostatic stellar evolution codes, the evaluation of the actual extension of this transition zone and the degree of mixing there would require more sophisticated tools. In any case, the larger the extension of this transition zone the deeper the resulting TDU (see, e.g., \cite{straniero2003}).\\
In addition to convection, other processes causing mixing below the convective envelope may affect the TDU and the formation of the $^{13}$C-pocket. Rotational induced instabilities were early considered \cite{herwig2003}, \cite{piersanti2013}. According to the extant models, mixing induced by rotation produces a marginal effect on the TDU, while it could modify the $^{13}$C-pocket, after its formation, and the consequent s-process nucleosynthesis. Nevertheless, recent asteroseismic studies of evolved low-mass stars revealed that most of the internal angular momentum is lost before the AGB phase, so that rotation likely does not play a relevant role in the AGB evolution and nucleosynthesis (see, e.g., \cite{moyano2022} and references therein). More promising is the hypothesis of mixing induced by internal gravity wave (IGW) generated at the boundary of the convective envelope \cite{den03}. The connection between internal convective zone and IGWs clearly emerges in various hydrodynamic simulations. This expectation is confirmed by the detection of g-mode (low-frequency) variability in photometry studies of main-sequence stars with convective cores \cite{bow19}. Likely, this process could also induce some mixing below the boundary of the convective envelope of AGB stars. The persistence of an internal magnetic field could also generate mixing in the He-rich zone, through the so-called magnetic buoyancy \cite{nor08}. \cite{ves21} have recently investigated the effect of this mechanism in low-mass AGB stars and conclude that the resulting s-process nucleosynthesis is in better agreement with the abundance patterns observed in AGB stars and with the isotopic composition of C-rich pre-solar grains that are supposed to originate in the cool atmosphere of C stars. However, the actual effect of all these (non-convective) mixing processes on the TDU effciency has not be clearly established yet.
\subsection{Predictions of extant AGB models}
\begin{figure}
\begin{center}
\includegraphics[scale=0.35]{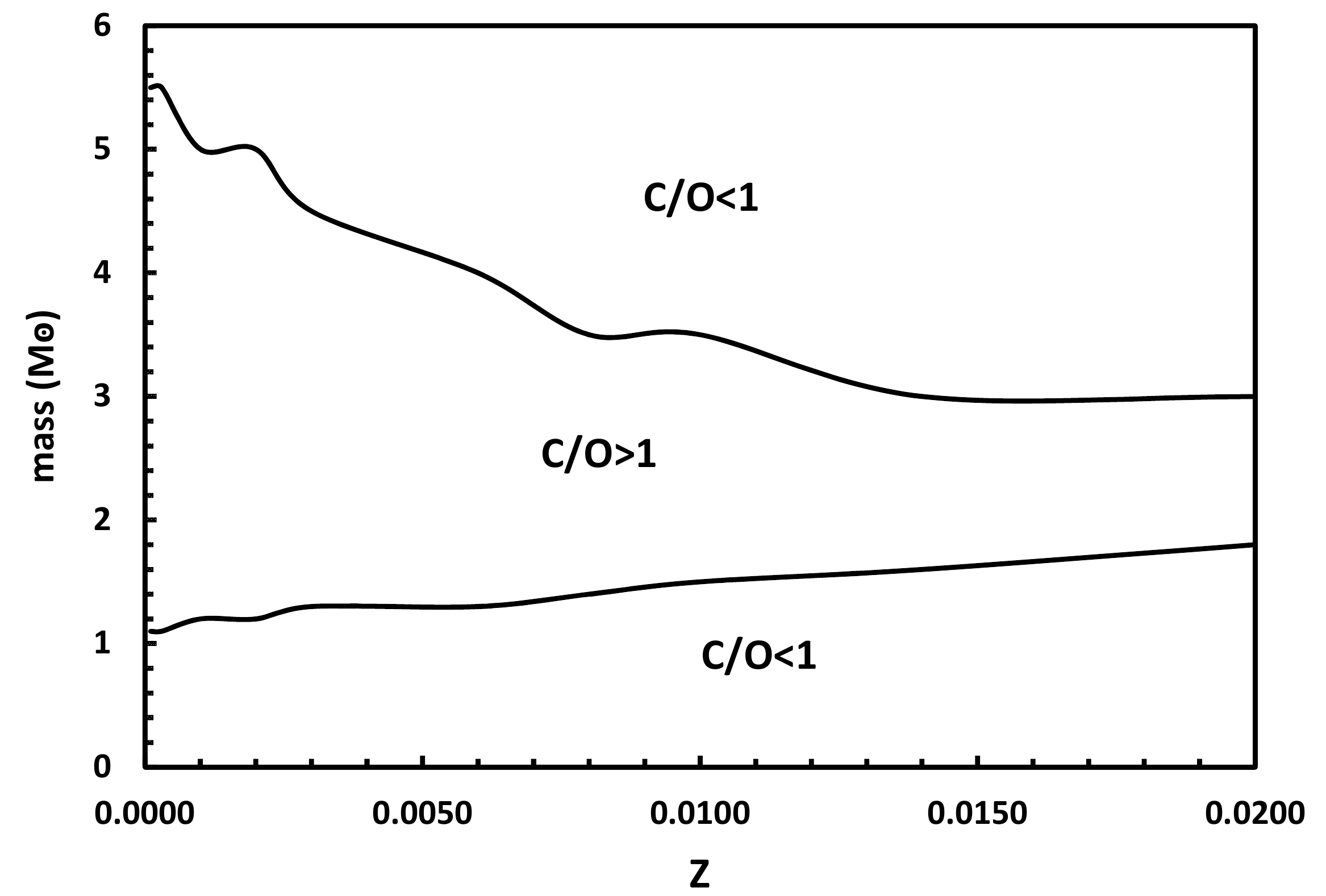}
\caption{Minimum and maximum initial mass for C-star progenitors versus metallicity, according to the theoretical predictions we obtain by means of the FuNS code (see text).
}\label{masse}
\end{center}
\end{figure}

Let us finally come back to to the C-star mystery. Although a reliable evaluation of which stars may become C-rich before leaving the AGB is still hampered by uncertainties on AGB mass loss and boundary mixing, extant stellar models provide a coherent, even if qualitative, picture. \\
In Figure \ref{masse}, we report the minimum and the maximum mass of stars expected to become C-rich during the AGB phase as a function of the metallicity. The minimum masses are from the FRUITY database (http://fruity.oa-teramo.inaf.it/) and were calculated by means of the FuNS code \cite{str06}. In particular, the AGB mass-loss rate was calculated by means of an empirical mass-loss vs period relation while the treatment of the boundary mixing is based on an exponential decay of the convective velocity below the convective envelope. The latter is a quite trivial consequence of the penetration of convective bubbles, which are accelerated in the convective envelope, into the underlying stable zone.     
When a bubble penetrate the stable zone, it is decelerated by the buoyancy at a rate:
\begin{equation}
    \ddot{r}=-\alpha v^2
\end{equation}
where $r$ is the distance from the internal border of the convective envelope and the $\alpha$ parameter has the dimension of the inverse of a distance. Hence, if $v_{0}$ is the velocity at $r=0$ (convective boundary), after a time integration, one may easily find: 
\begin{equation}
 \frac{v_0}{v}=\alpha v_0 t + 1
 \label{vvv}
\end{equation}
and, after a further integration:
\begin{equation}
 -r=\frac{1}{\alpha}\ln (\alpha v_0 t +1)
\end{equation}
finally, by means of equation \ref{vvv}:
\begin{equation}
 v=v_0\exp (-\alpha r)=v_0\exp (-\frac{r}{\beta H_P})
\end{equation}
where $\beta=1/\alpha H_P$ is a free dimensionless parameter and $H_P$ is the pressure scale-height. Note the similarity of the velocity gradient with the pressure gradient (except for the sign). Indeed, according to the hydrostatic equilibrium equation, the pressure gradient can be written as:
\begin{equation}
 P=P_0\exp (\frac{r}{H_P})
\end{equation}
(both r and P increase toward the centre). The resulting velocity and pressure within the convective boundary layer of a 2 M$_\odot$ model (solar initial composition), during a TDU episode, are shown in Figure \ref{velco} (arranged from \cite{str06}). This simple argument is confirmed by more sophisticated hydrodynamic simulations \cite{freytag2001}.
If some extra-mixing process is also at work, such as that due to IGWs or to magnetic buoyancy, the velocity profile within the boundary layer may be modified. In recent works, a convolution of two exponential decay functions is adopted in order to mimic this occurrence \cite{ves20}.

\begin{figure}
\begin{center}
\includegraphics[scale=0.35]{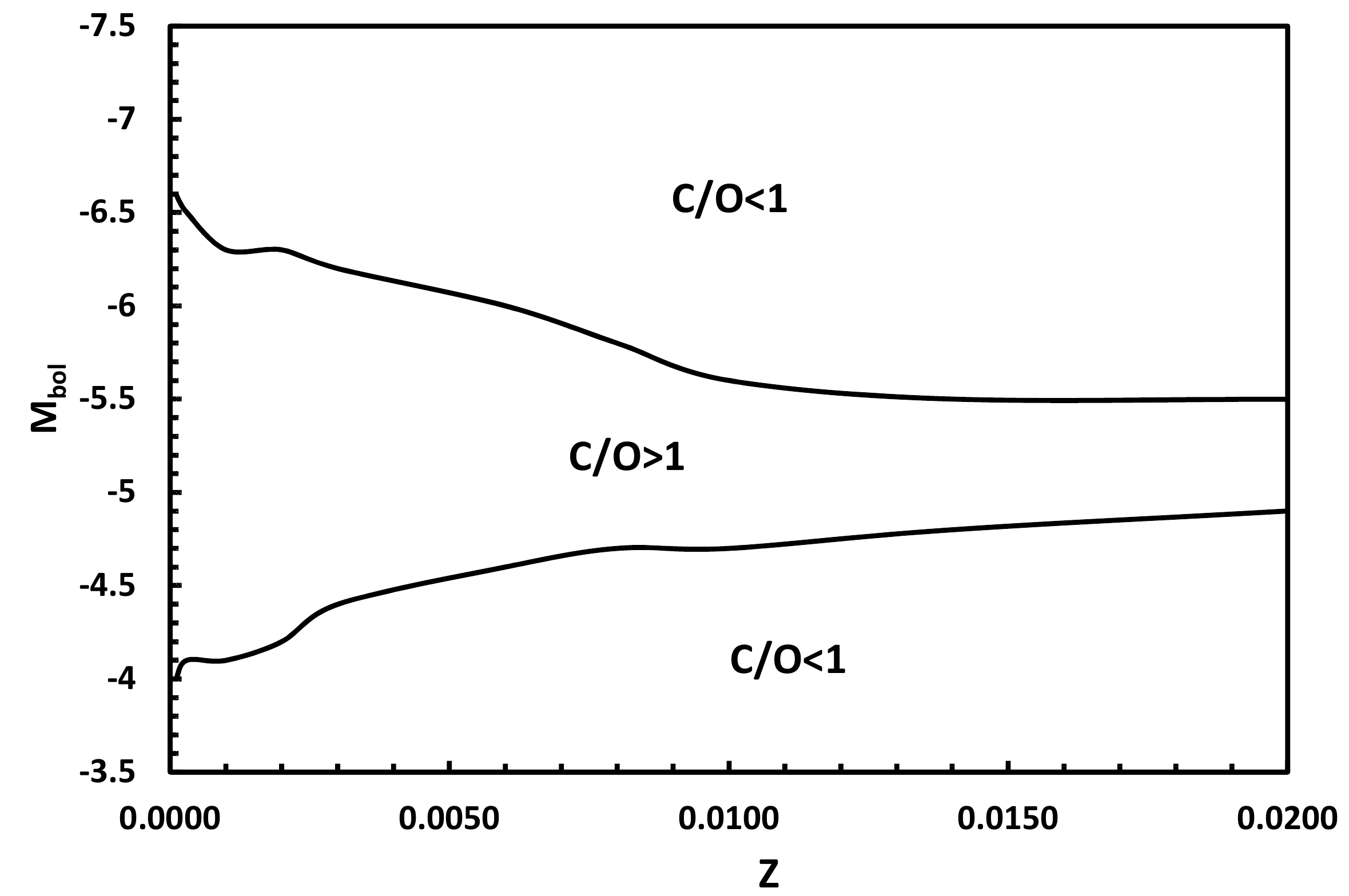}
\caption{Minimum and maximum bolometric magnitude for C stars versus metallicity, according to the theoretical predictions we obtain by means of the FuNS code (see text).
}\label{mbol}
\end{center}
\end{figure}
 The maximum masses in Figure \ref{masse} have been instead obtained by means of a more recent version of the FuNS code, in which a more appropriate numerical scheme to treat the HBB and HTDU is adopted. In particular, the differential equations describing the stellar structure in hydrostatic and thermal equilibrium are fully coupled to the differential equations describing the evolution of the internal composition, as due to mixing and thermonuclear burning processes. The qualitative picture is clear. The minimum mass is limited by the mass loss efficiency during both the pre-AGB and the AGB phases. It is lower at lower Z because less TDU episodes are needed before attaining the condition C/O$>1$. The maximum mass depends on the AGB mass-loss rate and it is limited by the onset of the HBB and the HTDU phenomena. In this case, a lower initial O abundance allows the formation of C stars with higher masses. 
Similarly, in Figure \ref{mbol} we show the corresponding maximum and minimum bolometric magnitude of C stars. Noteworthy, these  results coupled to the evolutionary time spend by each C-star progenitor up to the AGB phase and the duration of the C-star phase,  are the basic ingredients to construct theoretical C-star luminosity functions, but other ingredients are needed, such as the star formation history, the initial mass function and the metallicity vs age relation. As pointed out by \cite{sta10} (see also \cite{guandalini2013}), the luminosity function spread may be substantially affected by these additional parameters. 

\section{The observational framework}
Normal carbon stars represent a formidable challenge from the spectroscopic point
of view. They show very crowded spectra due to their
low temperatures (T$_{\rm{eff}}\sim 3000$ K) and strong molecular absorptions. Furthermore, most of AGB carbon stars are variable, thus their spectra are usually affected by large scale movements of the photosphere (stellar pulsations, shock waves...), which provoke strong line asymmetries, broadening and Doppler shifts. These phenomena 
greatly hinders the chemical analysis of these stars, and in principle would require the use of dynamical atmosphere models. Although some progress has been made in this sense \cite{eri14}, still most of the chemical analysis rely on the basis of static atmosphere models assuming LTE. This may introduce systematic errors in the determination of their chemical, and obviously, lead to wrong conclusions on the nucleosynthetic processes occurring in their interiors. Despite of this, considerable progress has been achieved in the past few decades and the abundance analyses show a comfortable agreement with theoretical predictions, in particular concerning the observed abundances of Li, F, C, N, O (and their isotopic ratios), and s-process elements. Next, we summarise the main achievements and the new issues that these abundance analyses have revealed. We will not discuss here the significant observational advances performed on the formation and structure of the circumstellar envelopes of carbon stars and their implications on the dust formation and the mass-loss rate history.\\
Carbon stars show only a few spectral windows (e.g. $\lambda\sim 4800-5000$ \AA, or $\lambda\sim 7700-8100$ \AA) suitable for abundance analysis at optical wavelengths provided that CN, C$_2$, CH and other C-bearing molecular features are included in any spectroscopic line list. In this sense, a significant effort has been made in the last years to improve the wavelength positions, energy levels and line intensities for all the isotopic combinations of the above mentioned molecules \cite{hed13}. The near infrared (NIR) spectrum of N-type stars
usually is less crowded allowing the identification of interesting atomic and molecular features, which makes the abundance analysis less difficult although still has to be explored with detail since many spectroscopic features (probably of atomic nature) are  unidentified \cite{nic17}. In any case, to perform an accurate abundance analysis in these stars the use very high resolution spectroscopy is mandatory, and unfortunately, still few high resolution NIR spectrographs attached to medium and/or large-size telescopes are available.

\subsection{The C/O and $^{12}$C/$^{13}$C ratios}
As mentioned in Sect. 2, carbon stars occupy the tip in
the AGB spectral sequence M→MS→S→SC→C(N),
thus they are the natural result of the continuous mixing of
carbon into the envelope throughout the TDU after each
TP. As a consequence the C/O ratio is expected to
increase continuously along the AGB phase until mass
loss terminates the evolution. Surprisingly, the derived C/O ratios so far do not greatly exceed unity ($\sim 1.0-1.5$, e.g. \cite{lam86}, \cite{abi02}). Only a few metal-poor N-type stars, observed in metal-poor extragalactic stellar systems, show significantly higher C/O ratios (4-8, \cite{del06}, \cite{abi08}). Although this is in agreement with the fact that the formation of a carbon star is easier at low metallicity because of the lower O content in the envelope and the increase of the efficiency of the TDU, the C/O ratios derived in the overwhelming majority  of carbon stars are still considerably lower than  theoretical predictions. 
Depending on the initial mass and metallicity, C/O ratios larger than $10$ are predicted at the end of the AGB. It has been suggested that the last part of the AGB evolution, is occupied by infrared extremely carbon-rich objects enshrouded in a thick dusty envelope, so that photospheric
abundances are not accessible. Also, as carbon exceeds oxygen in the envelope, it may condense into grains removing carbon atoms from the gas phase and, as consequence, keeping the C/O ratio only slightly larger than unity. In any case, this question remains unsolved.
Note that most of post-AGB stars show also C/O ratios very close to 1 \cite{kam22}, which is not either understandable on theoretical grounds.

Another issue directly related with the actual C/O ratio in the envelope of carbon stars is the $^{12}$C/$^{13}$C ratio. 
[34,51] identified a significant number ($\sim 25$ \%) of normal C stars with 
$^{12}$C/$^{13}$C$\sim 10-30$, while the average observed ratio is $\sim 70$ [33]; this latter value agrees with theoretical expectations when C/O exceeds unity in the envelope during the AGB phase. 
Indeed, this isotopic ratio is expected to increase continuously as $^{12}$C is being added into the envelope by the TDU to values larger than 100-200. Anomaly low carbon isotopic ratios are observed also in many low-mass RGB stars. Several authors (e.g. \cite{boo99}) have shown that this chemical anomaly in RGB stars may derive from transport mechanisms linking the envelope to zones where partial H-burning occurs. This phenomena is called “extra-mixing” or "deep mixing", and its 
nature is still highly debated \cite{kar10}. \cite{bus10} suggested that a similar mixing mechanism would operate
during the AGB in order to explain the low $^{12}$C/$^{13}$C values observed in some N-type stars. These authors show that, even assuming a normal $^{12}$C/$^{13}$C value ($\sim 20$) at the end of the RGB phase, ratios larger than $\sim 40$ are unavoidable reached when C/O$=1$ at the AGB phase. The operation of this mixing mechanism during the AGB phase would be, nevertheless, rather tricky since only for a few carbon stars $^{16}$O$/^{17}$O$/^{18}$O ratios compatible with the operation of such a mechanism are found, although their observed $^{12}$C/$^{13}$C and $^{14}$N/$^{15}$N ratios would be difficult to reconcile within this scenario \cite{hed13},\cite{abi17}. Note however, that  mainstream SiC grains, formed in the envelopes of N-type stars, provide the same evidence: in the St. Louis database, for 23\% of them, the carbon isotope ratio is below 40, while the average value is around 70, as observed in normal carbon stars. 

Observational evidence favouring  the existence of non-standard mixing mechanism(s) in the AGB phase is provided by the Li observations. 
Around  $2-3$ \% of galactic N-type stars show Li enhancements; a few show a huge 6708 \AA\- LiI absorption line; these stars are super Li-rich (A(Li)$\geq 4.0$, \cite{abi93b}). In fact AGB carbon stars are believed to be significant contributors to the Li budget in the Galaxy. Li can be produced in AGB stars by the operation of the Cameron \& Fowler mechanism \cite{cam71} at the bottom of a moderately hot (T$ >3.0 \times 10^7$ K) convective envelope. But, as mentioned in the previous section, these temperatures are reached in luminous (M$_{\rm{bol}}\leq -5.0$ mag) AGB stars with M$\geq 4-5$ M$_\odot$ where, in addition, the proton captures
on carbon at the bottom of the convective envelope may prevent the formation of a carbon star (see Fig. 2). In fact Li-enhancements are found in very luminous O-rich AGB stars \cite{gar07},\cite{ple93}. However, the fact that the luminosity function of N-type stars indicates that the overwhelming majority of them are of low-mass (M$\leq 3$ M$_\odot$) (see below), seems to discard the HBB as the mechanism responsible of the Li production in carbon stars. Note that some of the
Li-rich N-type stars are also $^{13}$C-rich. Similarly to the $^{12}$C/$^{13}$C issue, it has been suggested an explanation in terms of deep mixing \cite{pal11}. 

\subsection{Fluorine}
The source of fluorine in the Universe is currently widely debated,
and several sites have been proposed as potential candidates. However, only in AGB and post-AGB stars there is a direct
observation of fluorine production provided by spectroscopic
findings of photospheric [F/Fe]\footnote{We adopt the usual notation [X/Y]$\equiv$ log (X/Y) $-$ log (X/Y)$_\odot$ for the stellar value of any abundance ratio X/Y (by number). In the following we use ‘ls’ to refer to the light mass s-elements Y and Zr, and
‘hs’ to denote the high mass s-elements Ba, Nd, La and Sm.}  enhancements  \cite{wer05}, \cite{abi19}. Fluorine can be produced during the AGB phase through an intricate nuclear chain in such a way that its envelope abundance is expected to be correlated with the abundances of carbon and s-process elements. In fact this nuclear chain confers to this element both a primary and secondary origin. Recent F abundance determinations from HF lines at 2.3 $\mu$m in Galactic and extragalactic AGB carbon stars have confirmed large [F/Fe] enhancements as well as an increasing trend of this enhancement with the decreasing metallicity \cite{abi19}. However, while observations and theory agree at close-to-solar metallicity, stellar models at lower metallicities overestimate the fluorine production, in particular the abundance ratio between F and s-elements, which are also produced in AGB carbon stars. This discrepancy has lead to modify the driving process for the formation of the $^{13}$C-pocket with respect to the standard parameterisation (see Sect. 2). Recent AGB stellar models with mixing induced by magnetic buoyancy at the base of the convective envelope agree much better  with available fluorine spectroscopic measurements at low and close-to-solar metallicity \cite{ves21}. However, when the computed AGB fluorine yields are introduced in a galactic chemical evolution model, it becomes evident that other fluorine sources than AGB stars are required  \cite{gue19}.

\subsection{s-elements}
It was \cite{uts85} who firstly reported the enhancements of s-elements in the surface of carbon stars. This author found that N-type stars were typically of solar metallicity, presenting mean s-process element enhancements of a factor of 10 with respect to 
the Sun. However, more accurate studies, based on higher resolution spectra and better analysis tools \cite{lam86}, \cite{abi02}, \cite{del06}, \cite{abi01}, have led to strong revisions in the quantitative s-element abundances. N-type stars were confirmed to be of near solar metallicity, but they show on average $<$[ls/Fe]$>=+0.67\pm 0.10$ and $<$[hs/Fe]$>=+0.52\pm0.29$, which is significantly lower than that estimated
by \cite{uts85}. These values are of the same order as those derived in the O-rich S stars \cite{she18}. From these observations two main conclusions are reached: a) The abundance ratio between Rb and its neighbours (Sr, Y, Zr) indicates that the main neutron source operating in AGB stars is the $^{13}$C($\alpha,n)^{16}$O reaction and, therefore, than carbon stars are of low-mass ($\leq 3$ M$_\odot$); b) the [hs/ls] ratio (i.e. the abundance ratio between the heavy (Ba, La, Ce) and light (Sr, Y, Zr) s-elements), a parameter sensitive to the neutron exposure, increases with the decreasing metallicity of the star in agreement with theoretical predictions of the s-process. However, this ratio shows a significant dispersion at a given metallicity, which is usually interpreted as a signature of that existing in the $^{13}$C-pocket (abundance mass fraction profile, amount of $^{13}$C burnt) in AGB stars. This dispersion, on the other hand, is necessary to account for the s-element abundance patterns observed in individual C-stars.  We note, nevertheless, that the correlation  [hs/ls] vs. [Fe/H] is not clearly observed in post-AGB stars, the progeny of AGB stars, at least in the metallicity range studied \cite{kam22}. This shows the complexity of the s-process nucleosynthesis in AGB stars.

\subsection{Luminosity function}
Finally, the release of the Gaia DR3 catalogue made it possible to determine accurate
distances (and hence luminosities) to the Galactic AGB carbon  stars and constrain their positions in the HR diagram, their Galactic location and population membership. Figure 4 shows the luminosity function derived in a sample of $\sim 300$ Galactic carbon stars (N-type) with parallax accuracy better than $10\%$ according to Gaia DR3 \cite{abi22}. 
The average luminosity is M$_{\rm{bol}} = -5.04\pm 0.55$ mag, slightly
brighter than the average luminosity derived in carbon stars in the Magellanic Clouds. This figure agrees with theoretical expectations that C stars are formed more easily at low metallicities, thus earlier during the AGB phase (lower luminosity, see previous Section). However, Fig. 4 shows the existence of significant luminosity tails both at low and high M$_{\rm{bol}}$ values at which theoretically carbon stars would not exist because this would imply a progenitor mass low-er/hig\-her than the corresponding limits for their formation. Note however, as mentioned in Sect. 2, that other factors may affect the spread of the luminosity function. Nevertheless, the low luminosity tail (M$_{\rm{bol}}\geq -4.0$)  can be explained if a small fraction of the stars are extrinsic, i.e. they are low-mass stars ($<1.5$ M$_\odot$) that become C-rich because the accretion of carbon rich material in a binary system and then enter the AGB phase already with C/O$>1$ in the envelope. Other possibility is that the mass limit for the operation of efficient TDU and, thus, for the formation of a carbon star is lower than expected (see Fig. 2). Some observational evidence of this latter hypothesis exists \cite{she21}.  The high luminosity tail (M$_{\rm{bol}}\leq -5.5$) is more difficult to explain since these luminosities are attained by intermediate mass stars (M$>4-5$ M$_\odot$), which theoretically will not become a C star (at least with near solar metallicity, see Fig. 3) because the operation of the HBB. The existence of these high luminosity carbon stars (a few have been also observed in the Magellanic Clouds), implies that our understanding of the formation of a C star is still incomplete (see Sect. 2). 
\section{ A final remark}

There are also observational evidence of carbon enrichment occurring before the TP-AGB phase. These are the so called R-hot type carbon stars, with near solar metallicity and no s-element enhancement \cite{dominy1984,zamora2009}. Many of them have luminosities compatible  with red clump stars (central He burning) \cite{knapp2001,perryman1997}. The evolutionary phase that could explain the needed mixing may well be the He-flash, provided that He is ignited  at the border of the He-core, thus close to the H-shell, and at high degenerate physical conditions \cite{pac1977}. It was suggested that rotation may lead to this off-center ignition \cite{dominy1984,mengel1976}.  Moreover, as no R stars were found in binary systems, which statistically is unlikely, \cite{mcclure1997} suggested that they originate from binary mergers. 

No consistent evolutionary scenario has been found so far to explain these stars. One of the most popular, in terms of population synthesis analysis, is the merger of a He WD with a RGB star \cite{izzard2007}. After the merger, the resulting star evolves and, eventually, a degenerate He ignition occurs followed by a deep carbon dredge-up. \cite{piersanti2010} firstly investigated this scenario. Based on three-dimension SPH simulations of the merger of a He-WD with different masses and a RGB star, and one-dimension hydrostatic simulations of the accretion phase and the evolution up to the HB phase, they show that the dredge-up of freshly synthesised carbon does not occur in any of their  models.  This includes massive He WDs, that were found to be good candidates by \cite{zhang2013,zhang2020}.  In contrast, for massive He WDs, \cite{piersanti2010} obtained that if He is ignited after the accretion phase, the He-flash is mild  as the physical conditions at the border of the He-core are not highly degenerate. In this case, the convective He-shell remains confined inside the He-rich region and, later on, the entropy barrier due to the active H-burning shell, prevents any mixing.  On the other hand, if He ignites during the accretion phase, the accretion disk prevents any penetration of the convective envelope. However, all these calculations of He flash models were done with a one dimension hydrostatic code and this occurrence may likely be their major limit. \cite{mocak2010} argue that to properly treat convection, a three-dimension hydrodynamical model should be preferred. However, this type of numerical simulations only covers about 1 day of the stellar evolution. The models performed by \cite{mocak2010} show the growth of the He-convective unstable zone toward the H-rich layers on a dynamical timescale; some mixing could occur, but it does not take place during the simulation. Thus, the question remains open.

Recently, by analysing a large sample of Galactic carbon stars for which very accurate  astrometry is available from Gaia DR3, [46] found many R-hot stars with low luminosities, covering all the RGB phase. This shows that, at least for some R-hot stars, carbon enrichment should occur before the He-flash. In this framework, an extrinsic origin appears favoured.

\section{Future perspectives}
A bright future is expected for studies on both normal and R-type C stars. The detection of isotopic abundances is a key tool to understand the interplay between mixing and burning processes in stars. However, the optical and near-IR spectrum of a C star is a forest of closely-spaced absorption lines. Thus, very high spectral resolution, coupled to a high signal-to-noise ratio, is required to distinguish lines of different isotopes. As a matter of fact, only some isotopic abundances of C and O have been measured so far. In this context, the next generation of large aperture telescopes (ELT and its competitors) will provide a great opportunity to improve our understanding of C stars. For instance, the high-resolution ELT instrument ANDES, formerly known as HIRES, will deliver high-quality stellar spectra (R$> 100,000$ and S/N$> 100$) of AGB stars belonging to the Local Group of Galaxies \cite{hires2013}. This observational effort must be necessarily accompanied by the identification and improvement of the spectroscopic parameters of C- and O-bearing molecular lines in the visual and infrared wavelength ranges.\\
On the other hand, asteroseismic studies can provide an unprecedented view on the internal structure of stars in different phases  and their evolution. So far, data from CoRoT, Kepler, K2, and now TESS, have demonstrated the relevance and potential of this
novel technique in understanding stellar physics. The detection of frequencies of p and g-modes allows us to infer both average and localised properties of different internal regions of the target stars. In particular, the internal rotation rate may be measured, and important phenomena, such as gravity waves and magnetic buoyancy, may be constrained. In addition, the chemical composition gradient, the thermal stratification and the sound speed profile could be also deduced from frequency patterns. The next ESA missions, PLATO and, hopefully, HAYDIN, will deliver more accurate data to improve our understanding of these processes.  
 
Finally, we can easily guess that all these future studies, thanks to the superior quality of the observational data, will motivate and boost the development of new and more sophisticated stellar models, capable to accurately describe the non-standard processes involved in the formation and evolution of intrinsic C star. 

\begin{figure}
  \includegraphics[width=8.0 cm]{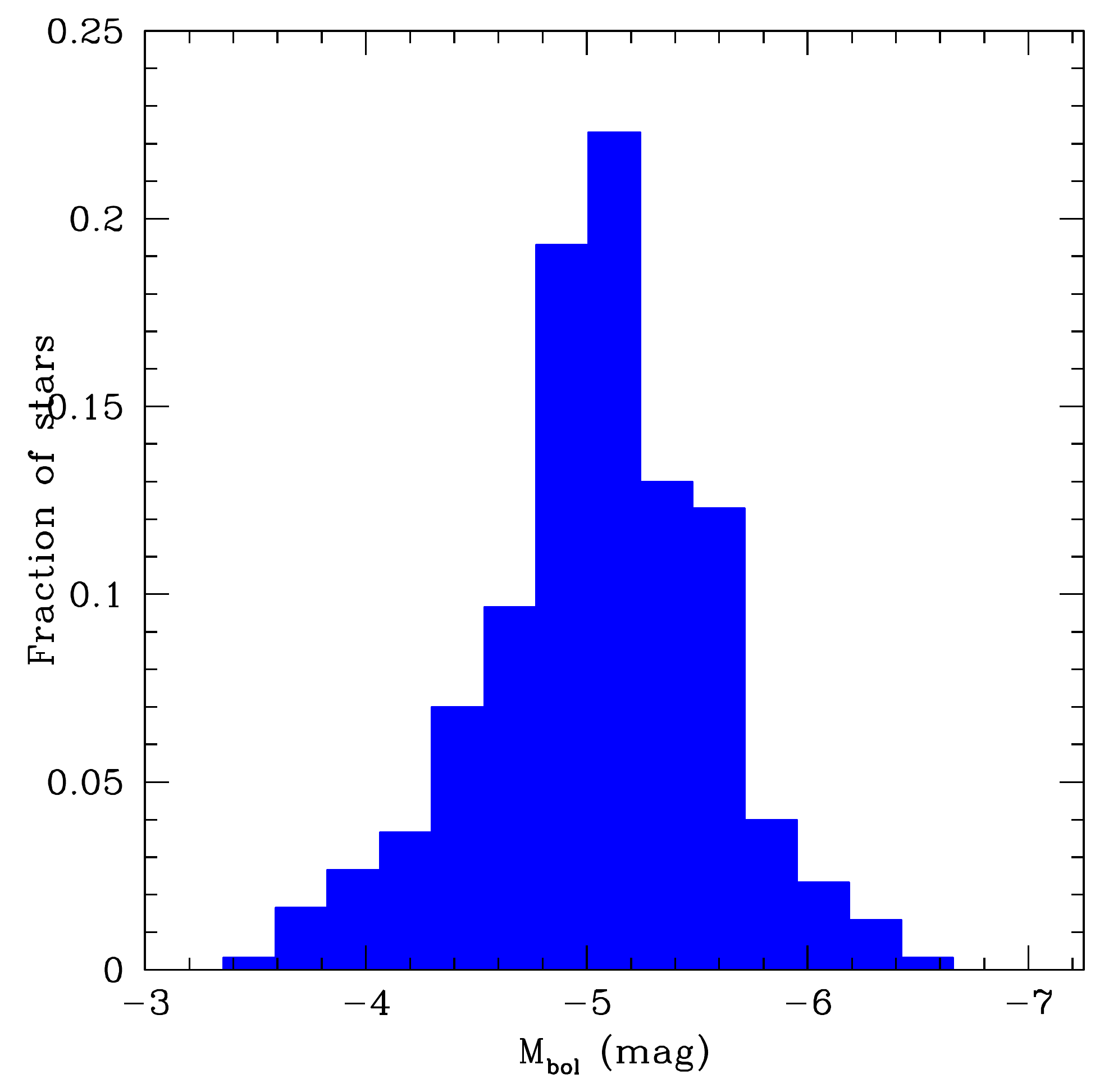}
\caption{Luminosity function for about 300 Galactic AGB carbon stars derived from Gaia DR3 parallaxes with uncertainty less than 10\%. The bin size is 0.25 mag (see text). Adapted from \cite{abi22}.}
\label{fig:1}       
\end{figure}

\begin{acknowledgements}
This work has been supported by the Spanish project PGC2018-095317-B-C21 financed by the MCIN \\/\-AEI FEDER “Una manera de hacer Europa”, and by the project PID2021-123110NB-I00 financed by MCIN/AEI \\ /10.13039/501100011033/FEDER, UE. 

\end{acknowledgements}

\bibliographystyle{spphys}       
\bibliography{cstar}   

%
%
\end{document}